\documentclass[twocolumn,aps,prl,superscriptaddress]{revtex4}
\usepackage[plainpages=false,pdfpagelabels,colorlinks=true,linkcolor=red,urlcolor=blue,citecolor=blue,pdftitle={Title},pdfauthor={},pdfdisplaydoctitle=true,pdfduplex=DuplexFlipLongEdge]{hyperref}
\usepackage{verbatim}
\usepackage{amsmath,amssymb,bm}
\usepackage{amsfonts}
\usepackage{braket}
\usepackage{float}
\usepackage{subfig}
\usepackage{graphicx,xfrac,appendix}
\usepackage[english]{babel}
\makeatletter
\@ifundefined{textcolor}{}
{%
 \definecolor{BLACK}{gray}{0}
 \definecolor{WHITE}{gray}{1}
 \definecolor{RED}{rgb}{1,0,0}
 \definecolor{GREEN}{rgb}{0,1,0}
 \definecolor{BLUE}{rgb}{0,0,1}
 \definecolor{CYAN}{cmyk}{1,0,0,0}
 \definecolor{MAGENTA}{cmyk}{0,1,0,0}
 \definecolor{YELLOW}{cmyk}{0,0,1,0}
}

\usepackage{epstopdf}
\usepackage{xcolor}
\usepackage{tikz}

\makeatother

\begin{document}
\title{Large Chern numbers in a dissipative dice model }

\author{Shujie Cheng}
\affiliation{Department of Physics, Zhejiang Normal University, Jinhua 321004, China}
\author{Gao Xianlong}
\affiliation{Department of Physics, Zhejiang Normal University, Jinhua 321004, China}
\date{\today}

\begin{abstract}
 For decades, the topological phenomena in quantum systems have always been catching our attention. Recently, there are many interests on the systems where topologically protected edge states exist, even in the presence of non-Hermiticity. Motivated by these researches, the topological properties of a non-Hermitian dice model are studied in two non-Hermitian cases, viz. in the imbalanced and the balanced dissipations.
 Our results suggest that the topological phases are
 protected by the real gaps and the bulk-edge correspondence readily seen in the real edge-state spectra.
 Besides, we show that the principle of the bulk-edge correspondence in Hermitian case is still effective in analyzing the three-band
 non-Hermitian system. We find that there are topological non-trivial phases with large Chern numbers $C=-3$ robust against the dissipative perturbations.
 \end{abstract}

\maketitle

\section{Introduction}
Topology acts as a critical role in physics, and the topological phenomena have attracted increasing interests
in the fields of the topological condensed matters \cite{condensed-matter_1,condensed-matter_2}, topological photonics systems \cite{photonics_5,photonics_6}, and
the ultracold atomic systems \cite{ultracold_6,ultracold_8,ultracold_9},
in which the characterization of topological properties is out of the framework of spontaneous symmetry breaking \cite{SSB_1,SSB_2}.
Besides, due to the symmetries, the topological classification of Hermitian systems was well established \cite{H_c1,H_c2}. A typical
signature of topology is the bulk-edge correspondence, which typically gives the relationships between the topological invariants and the associated
edge states \cite{bulk-edge_1,bulk-edge_2} and makes insulators \cite{condensed-matter_1,insulator_1,insulator_2,insulator_3,insulator_4,insulator_5,insulator_6,insulator_7} and superconductors \cite{condensed-matter_2,superconductor_1,superconductor_2,superconductor_3} fantastic and applicable.

Beyond the investigated topological phenomena in Hermitian systems, there are great attention on the framework of the open systems, in which the energy or particles are no longer conserved \cite{open_1,open_2}. Generally speaking, open systems are also interpreted as the non-Hermitian systems described by the effective non-Hermitian Hamiltonians. Recent years there are plenty of progress made in theoretical \cite{theory_01,theory_02,theory_03,theory_04,theory_05,theory_1,theory_2,theory_3,theory_4,theory_5,theory_7,theory_8,theory_9_1,theory_9_2,theory_10_1,theory_10_2,theory_11,theory_12,theory_13_1,theory_13_2,theory_13_3,theory_14,theory_15,theory_16,theory_17,theory_18,theory_19,theory_20,theory_21,theory_22,theory_24,theory_25} and experimental \cite{exper_1,exper_2,exper_3,exper_4,exper_5,exper_6,exper_7,exper_8,exper_9} researches in revealing the topological nature of the systems in the presence of non-Hermiticity. Different from the Hermitian systems, it was studied that there were 38-fold symmetries in the non-Hermitian case \cite{theory_9_1},
leading to the well-established topological classification of the non-Hermitian systems.

Similar to the Hermitian case, the bulk-edge correspondence remains a central topic in the non-Hermitian cases. However,
there is a subtle issue on how the bulk-edge correspondence exists. Previous works show that the bulk-edge correspondence
maintains by the Bloch topological invariants \cite{theory_1,theory_2,theory_3,theory_4,theory_5,theory_7} in the presence of
the non-Hermiticity. Moreover, the non-Hermiticity can lead to anomalous edge states \cite{theory_8}. Furthermore, the non-Hermiticity can be induced by the imbalanced tunnelings, leading to the non-Hermitian skin effect \cite{theory_10_1,theory_10_2}.
In such a system, the conventional bulk-edge correspondence is altered and replaced by the generalized bulk-edge correspondence suggested by the non-Bloch band theory \cite{theory_10_1,theory_10_2}.

Combined with the recent researches on Dice models \cite{dice_1,dice_2,dice_3}, in which large Chern numbers and multiple edge states are
uncovered, in this paper, we are motivated to study the topological properties of a dissipative non-Hermitian Dice model.
To be concrete, we make attempt to answer the questions whether large Chern numbers are robust against  the dissipative perturbations and how the
bulk-edge correspondence exists.

The rest of paper are organized as follows. Section \ref{S2} describes our general non-Hermitian model. Section \ref{S3} briefly
makes a definition of the Chern numbers of the non-Hermitian systems and specifies two Chern markers to characterize the
topological properties of systems. Section \ref{S4} involves the analyses and discussions of two typically dissipative cases.
Finally, we make a conclusion in Sec.~\ref{S5}.

\section{\label{S2}Model and Hamiltonian}
\begin{figure}[H]
\centering
\includegraphics[width=0.5\textwidth]{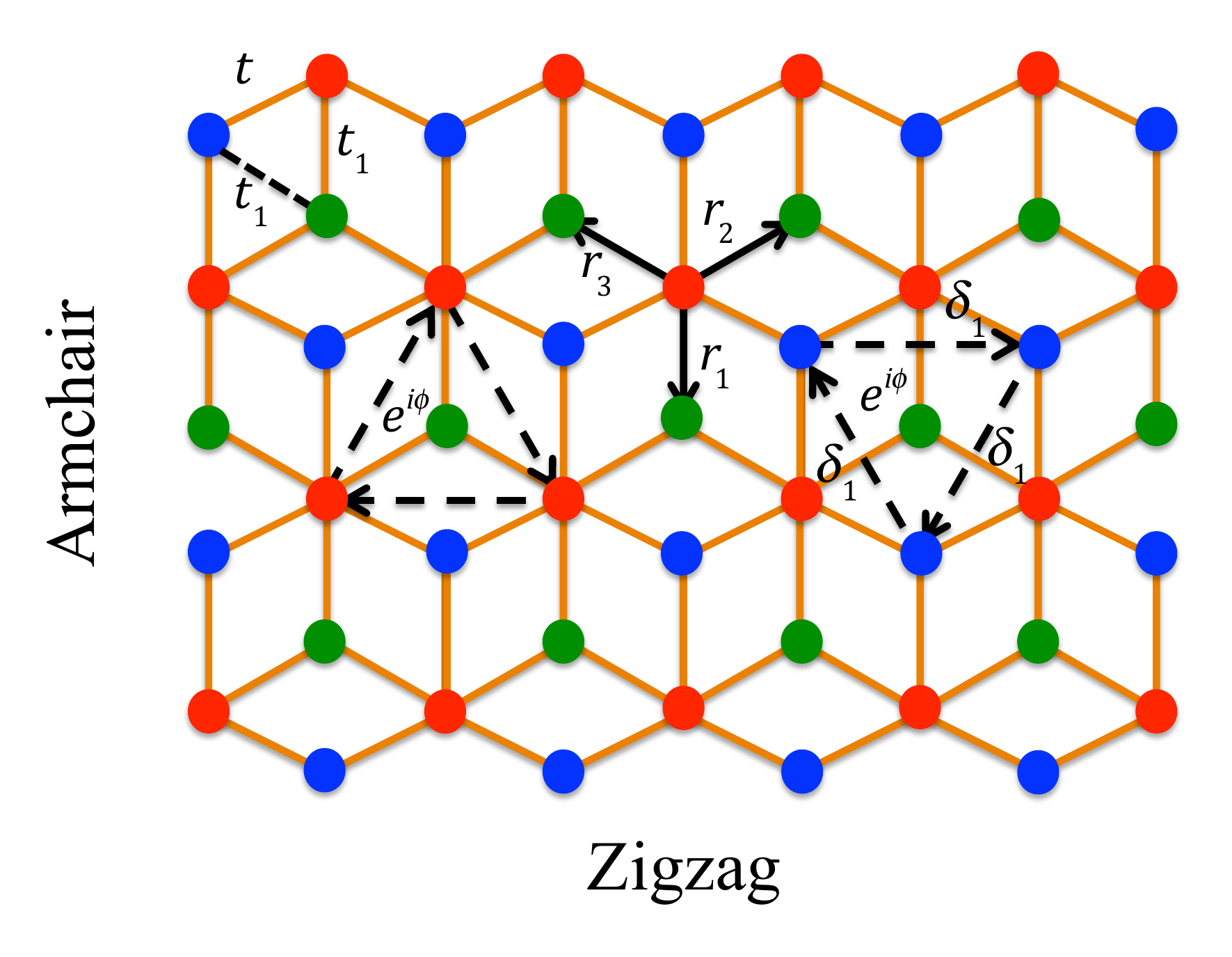}
\caption{The sketch of dice lattice formed by staggered arrangement of sublattice R (red), B (blue), and G (green).
The Hamiltonian $\hat{H}$ consists of the nearest neighbor tunnelings with hopping amplitude $t$ between one R site
and one B site, and $t_{1}$ between one G site and one R or B site, and the phase-dependent
next-nearest-neighbor tunnelings with hopping amplitude $t_{2}e^{i\phi}$ between the same R or B sublattice sites.
Moreover, each lattice site is subject to the potentials and dissipations. The vectors $\mathbf{r}_{s}$ (s=1,2,3) point from the centering site
to its neighboring sites and the vectors $\mathbf{\delta}_{s}$ (s=1,2,3) connect the next-nearest-neighbor R or B sites. Zigzag
and armchair are two common geometric boundary shapes, which will be used for the discussions of the
bulk-edge correspondence.
 }
\label{f1}
\end{figure}

In this paper, we study a non-interacting model with dissipative on-site potentials based on a dice lattice \cite{dicelattice_1,dicelattice_2,dicelattice_3,dicelattice_4,dicelattice_5}, shown in Fig.~\ref{f1}. Intuitively, there
are three non-equivalent sublattice sites, marked as R (red dots), B (blue dots), and G (green dots). Thus the lattice model belongs to the SU(3) system \cite{dice_1,dice_2,dice_3,ultracold_2,SU3}. The single-particle Hamiltonian consists of the following two parts,
\begin{equation}\label{eq1}
\hat{H}=\hat{H}_{1}+\hat{H}_{2}.
\end{equation}

$\hat{H}_{1}$ denotes the Hermitian part, which has been proposed in Ref.~\cite{dice_2} with
 \begin{equation}\label{eq2}
 \begin{aligned}
 &\hat{H}_{1}=\left[\sum_{\langle\mathbf{R}_i,\mathbf{B}_j\rangle}t\hat{c}^\dag_{\mathbf{R}_i}\hat{c}_{\mathbf{B}_j}
 +\sum_{\langle\mathbf{R}_i,\mathbf{G}_\ell\rangle}t_1\hat{c}^\dag_{\mathbf{R}_i}\hat{c}_{\mathbf{G}_\ell}\right.\\
 &\left.+\sum_{\langle\mathbf{B}_j,\mathbf{G}_\ell\rangle}t_1\hat{c}^\dag_{\mathbf{B}_j}\hat{c}_{\mathbf{G}_\ell}+
 \sum_{\langle\mathbf{R}_i,\mathbf{R}_j\rangle}t_2e^{i\phi}
 \hat{c}^\dag_{\mathbf{R}_i}\hat{c}_{\mathbf{R}_j}\right.\\
 &\left.+\sum_{\langle\mathbf{B}_i,\mathbf{B}_j\rangle}t_2e^{i\phi}
 \hat{c}^\dag_{\mathbf{B}_i}\hat{c}_{\mathbf{B}_j}+H.c.\right]+\gamma_{1}\Delta\sum_{\mathbf{R}_i}\hat{c}^\dag_{\mathbf{R}_i}\hat{c}_{\mathbf{R}_i}\\
 &+\gamma_{1}\Delta\sum_{\mathbf{B}_i}
  \hat{c}^\dag_{\mathbf{B}_i}\hat{c}_{\mathbf{B}_i}+\gamma_{2}\Delta\sum_{\mathbf{G}_i}\hat{c}^\dag_{\mathbf{G}_i}\hat{c}_{\mathbf{G}_i},
 \end{aligned}
 \end{equation}
 in which $t$ is the hopping amplitude between one $R$ site and one $B$ site, $t_{1}$ is the hopping amplitude between one $G$ site and one
 $R$ or $B$ site, $t_{2}e^{-i\phi}$ is the complex tunneling amplitude of the next-nearest-neighbor tunnelings between the same adjacent
 sublattices with $\phi$ being the phase, and $\gamma_{1}\Delta$ and $\gamma_{2}\Delta$ denote strengths of on-site potentials with $\Delta$
 being the modulation parameter and $\gamma_{1}$ and $\gamma_{2}$ the modulation rate of $\Delta$.

 $\hat{H}_{2}$ represents the non-Hermitian part which shows that each sublattice is attached with dissipation, and is proposed as
 \begin{equation}\label{eq3}
 \hat{H}_{2}=i\eta\sum_{\mathbf{R}_i}\hat{c}^\dag_{\mathbf{R}_i}\hat{c}_{\mathbf{R}_i}+
 i\eta\sum_{\mathbf{B}_i}\hat{c}^\dag_{\mathbf{B}_i}\hat{c}_{\mathbf{B}_i}+i\chi\eta_{1}\sum_{\mathbf{G}_i}\hat{c}^\dag_{\mathbf{G}_i}\hat{c}_{\mathbf{G}_i},
 \end{equation}
 where $\eta$ and $\eta_{1}$ are the strengths of the dissipations with the same sign, and $\chi\in\pm 1$ plays the role of a switch by weighing overall
 gain and loss of the system. When $\eta=\eta_{1}=0$, the system becomes the Hermitian dice model \cite{dice_2}, and if the tunnelings between
 G sites and B sites are omitted, then it goes back to the Haldane-like Dice model \cite{dice_3}.

Under the discrete translational symmetry, the Hamiltonian $\hat{H}_{1}$ can be mapped into momentum space by using the discrete Fourier transformation
 \begin{equation}\label{eq4}
 \hat{c}_{\mathbf{k}, \alpha}=\frac{1}{\sqrt{N}} \sum_{{\mathbf \alpha}_{j}} e^{-i \mathbf{k} \cdot {\mathbf\alpha}_{j}} \hat{c}_{{\mathbf \alpha}_j},
 \end{equation}
 where $N$ is the total number of the unit cells, and $\alpha\in\{\rm R, B, G\}$ denotes the type of sublattice with $\mathbf{\alpha}_{j}$ being the corresponding
 coordinate. Thus, the Bloch Hamiltonian has a form as
 \begin{equation}\label{eq5}
 \hat{H}(\mathbf{k})=\sum_{\mathbf{k}}\hat{c}^{\dag}_{\mathbf{k}} \mathcal{\hat{H}}(\mathbf{k}) \hat{c}_{\mathbf{k}},
 \end{equation}
 where $\hat{c}^{\dag}_{\mathbf{k}}=\left(\hat{c}_{\mathbf{k}, \rm R}, \hat{c}_{\mathbf{k}, \rm B}, \hat{c}_{\mathbf{k}, \rm G}\right)^{T}$ is the three-component
 basis, and $\mathcal{H}(k)$ is expressed as
 \begin{equation}\label{eq6}
 \mathcal{H}_{\mathbf k}=I(\mathbf{k},\eta,\eta_{1})+\mathbf{d(k)}\cdot\vec{\lambda},
 \end{equation}
 where $I(\mathbf{k},\eta,\eta_{1})$ is the scalar leading to a overall shift of the energy in a complex space, $\mathbf{d(k)}$ is a coefficient vector, and
 $\vec{\lambda}$ is a vector consisting of the Gell-Mann matrices \cite{SU3,ultracold_2,dice_2}. Concretely, the components of $\mathbf{d(k)}$ are displayed as
 \begin{equation}\label{eq7}
 \begin{aligned}
 d_1&=t\sum_{s}\cos\left(\mathbf{k}\cdot\bm{r}_s\right),~~d_2=t\sum_{s}\sin\left(\mathbf{k}\cdot\bm{r}_s\right),\\
 d_4&=d_6=t_1\sum_{s}\cos\left(\mathbf{k}\cdot\bm{r}_s\right),\\
 d_7&=-d_5=t_1\sum_{s}\sin\left(\mathbf{k}\cdot\bm{r}_s\right),\\
 d_3&=-2t_2\sin{\phi}\sum_{s}\sin(\mathbf{k}\cdot\bm{\delta}_s)+\frac{i\eta}{3},\\
 d_8&=\frac{\gamma_{1}-\gamma_{2}}{\sqrt{3}}\Delta+\frac{2t_2}{\sqrt{3}}\cos{\phi}\sum_{s}\cos(\mathbf{k}\cdot\bm{\delta}_s)+\frac{i(2\eta-\chi\eta_{1})}{\sqrt{3}},
 \end{aligned}
 \end{equation}
 with six vectors $\bm{r}_{s}$ and $\bm{\delta}_{s}$ (s=1,2,3) being expressed as
 \begin{equation}\label{eq8}
 \begin{aligned}
 \bm{r}_{1}&=\dbinom{0}{-1},~\bm{r}_{2}=\frac{1}{2}\dbinom{\sqrt{3}}{1},~\bm{r}_{3}=\frac{1}{2}\dbinom{-\sqrt{3}}{1},\\
 \bm{\delta}_{1}&=\dbinom{\sqrt{3}}{0},~\bm{\delta}_{2}=\frac{1}{2}\dbinom{-\sqrt{3}}{3},~\bm{\delta}_{3}=-\frac{1}{2}\dbinom{\sqrt{3}}{3}.
 \end{aligned}
 \end{equation}

\section{\label{S3} Chern numbers}
According to the 38-fold topological classification for the non-Hermitian systems, our systems under research belong to the A class
in complex Altland-Zirnbauer symmetry \cite{theory_9_1}. In this section, we briefly describe how to define the Chern numbers of associated energy bands of
our systems. To begin with, we discuss the eigenstates of the non-Hermitian systems. Compared with the Hermitian systems, the eigenstates in the non-Hermitian systems form the biorthogonal bases \cite{theory_9_2}, which satisfy
 \begin{equation}\label{eq9}
 \begin{aligned}
 \mathcal{H}_{\mathbf k}\ket{\psi^{R}_{n}}&=E_{n}\ket{\psi^{R}_{n}},\\
 \mathcal{H}^{\dag}_{\mathbf k}\ket{\psi^{L}_{n}}&=E^{*}_{n}\ket{\psi^{L}_{n}},\\
 \end{aligned}
 \end{equation}
where $\ket{\psi^{R}_{n}}$ denotes the right eigenstate of $\mathcal{H}_{\mathbf k}$ and $\ket{\psi^{L}_{n}}$ is the left eigenstate.
We define the bottom band, middle band and top band according to the real eigenvalues of $\mathcal{H}_{\mathbf k}$, and the ascending
value of $n$ ($n=1,2,3$) corresponds to the three bands from bottom to top. Furthermore, the biorthogonal bases obey such a biorthogonality relation
\begin{equation}\label{eq10}
\braket{\psi^{R}_{n'}|\psi^{L}_{n}}=\braket{\psi^{L}_{n'}|\psi^{R}_{n}}=\delta_{n'n},
\end{equation}
and fulfill the completeness condition
\begin{equation}\label{eq11}
\sum_{n}\ket{\psi^{L}_{n}}\bra{\psi^{R}_{n}}=\sum_{n}\ket{\psi^{R}_{n}}\bra{\psi^{L}_{n}}=I.
\end{equation}

With these definitions, the  $\mathbb{Z}$ topological invariant, namely the Chern number is defined as \cite{theory_10_1,theory_13_1,theory_13_2,theory_13_3}
\begin{equation}\label{eq12}
C_{n}=\frac{1}{2\pi}\int_{\rm{1st}~BZ} \Omega_{n}(\mathbf{k})d^2\mathbf{k},
\end{equation}
in which $\Omega_{n}(\mathbf{k})=\partial_{k_{x}}A^{y}_{n}(\mathbf{k})-\partial_{k_{y}}A^{x}_{n}(\mathbf{k})$ denotes the Berry curvature of the nth band
and $A^{j}_{n}(\mathbf{k})=i\braket{\psi^{L}_{n}(\mathbf{k})|\partial_{k_{j}}\psi^{R}_{n}(\mathbf{k})}$ ($j=x,y$) is the Berry connection, obtained
from the non-Hermitian Hamiltonian. Hereinbelow, for convenience, we will use two Chern markers, namely $C_{\frac{1}{3}}=C_{1}$ and $C_{\frac{2}{3}}=C_{1}+C_{2}$
to characterize the topological phases of the non-Hermitian systems.

\section{\label{S4} Results and Discussions}
In this section, two non-Hermitian cases will be investigated, i.e., the imbalanced dissipation and the balanced dissipation, respectively.
The imbalanced dissipation denotes that the overall gain and loss are not conserved and the balanced dissipation refers to the case
where the gain and loss are balanced. Without loss of generality, we fix $t$ as the unit of energy and set some other global parameters
with $t_1=0.5t$, $t_{2}=0.526t$, $\phi=\frac{\pi}{2}$, $\gamma_{1}=5$, and $\gamma_{2}=7$. The following main effort is focused
on the investigation in the large Chern numbers and the bulk-edge correspondence with the existence of dissipations.

\begin{figure}[H]
\centering
\includegraphics[width=0.5\textwidth]{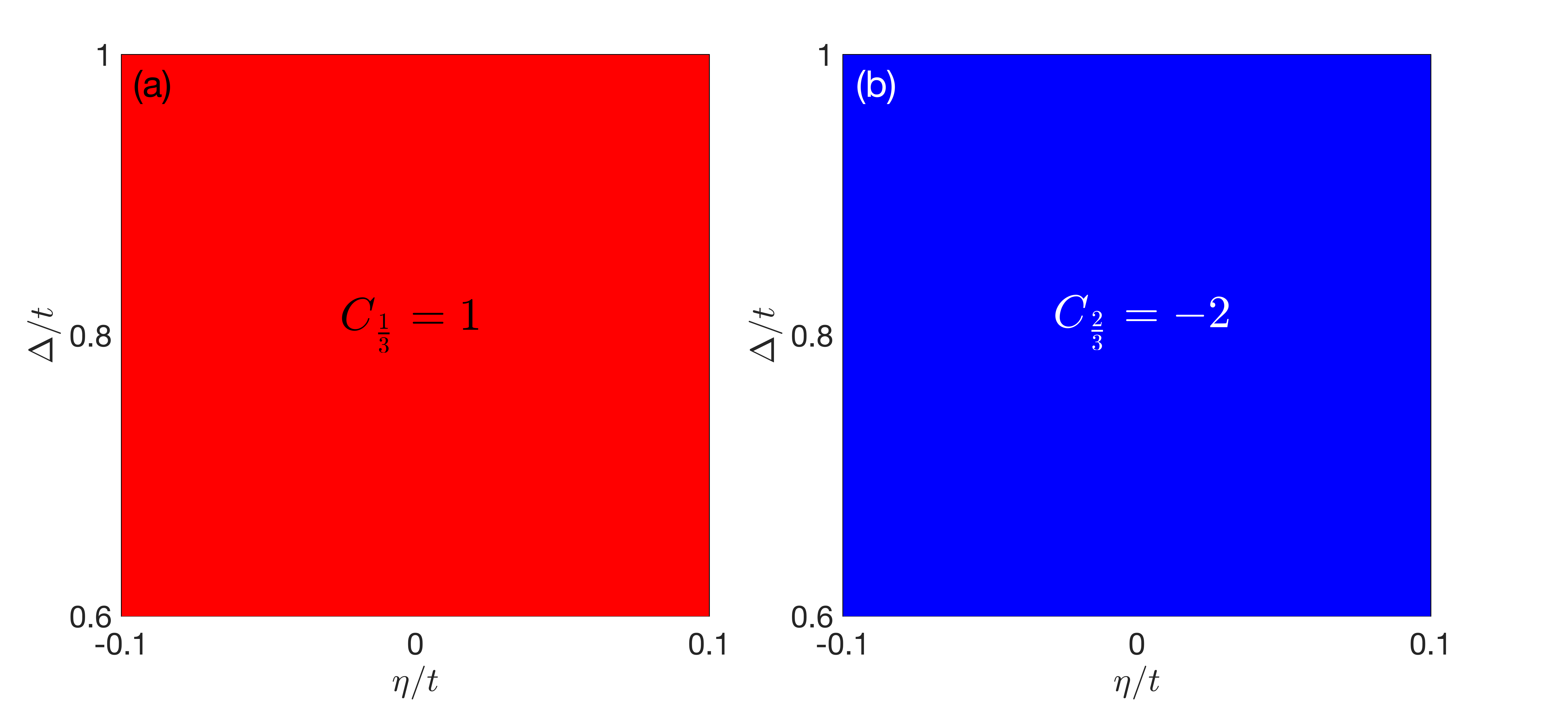}
\caption{ Two phase diagrams in the $\Delta$-$\eta$ parameter space. (a) $C_{\frac{1}{3}}=1$. There is a topological
non-trivial bottom band with Chern number $C_{1}=C_{\frac{1}{3}}=1$. (b) $C_{\frac{2}{3}}$=-2. There is a topological non-trivial
middle band with large Chern number $C_{2}=C_{\frac{2}{3}}-C_{\frac{1}{3}}=-3$. The involved parameters are $\Delta\in[0.6t,1t]$,
$\eta=\eta_{1}\in[-0.1t,0.1t]$, and $\chi=1$.}
\label{f2}
\end{figure}

\subsection{Imbalanced dissipation}
By taking $\eta_{1}=2\eta$ ($\eta\in[-0.1t, 0.1t]$) and $\chi=1$, the overall gain and loss are not conserved. Furthermore, the value of $\eta$ determines whether the non-Hermitian potential is pure gain or loss type. And as a result, the system belongs to the imbalanced dissipation case. With these matrix elements in Eq.~(\ref{eq7}),
we calculate the Chern numbers by means of the definition in Eq.~(\ref{eq12}) and two phase diagrams are plotted
in Figs.~\ref{f2}(a) and \ref{f2}(b), respectively. From Fig.~\ref{f2}(a), we know that in the chosen parameter region,
the bottom band of system is topological nontrivial with Chern number $C_{1}=C_{\frac{1}{3}}=1$. From Fig.~\ref{f2}(b),
we intuitively notice that there is a large Chern number of the middle band of the system. Through a simple mathematical
relationship, the middle band has a large Chern number with $C_{2}=C_{\frac{2}{3}}-C_{1}=-3$. In addition, the topological
properties of bands are protected by real gaps \cite{theory_1,theory_2}. To check this point, we then choose three parameter
points ($\eta_{a}, \Delta_{a}$)=(-0.1t, 0.6t), ($\eta_{b}$, $\Delta_{b}$)=(0.1t, 0.8t), and ($\eta_{c}, \Delta_{c}$)=(0.1t, 1t) in the phase diagram.
We here point out that we have tested that the parameter interval supporting non-zero Chern number is relatively large, which is quite beneficial to make use of the topological properties of the system. However, when the dissipation strength further increases, the real energy spectra becomes gapless, leading to the result that fails to define an effective Chern number \cite{theory_10_1,theory_13_1,theory_13_2,theory_13_3}. We here in the paper concentrate on the cases with the dissipation strength where the real energy spectra are gapped.

\begin{figure}[H]
\centering
\includegraphics[width=0.5\textwidth]{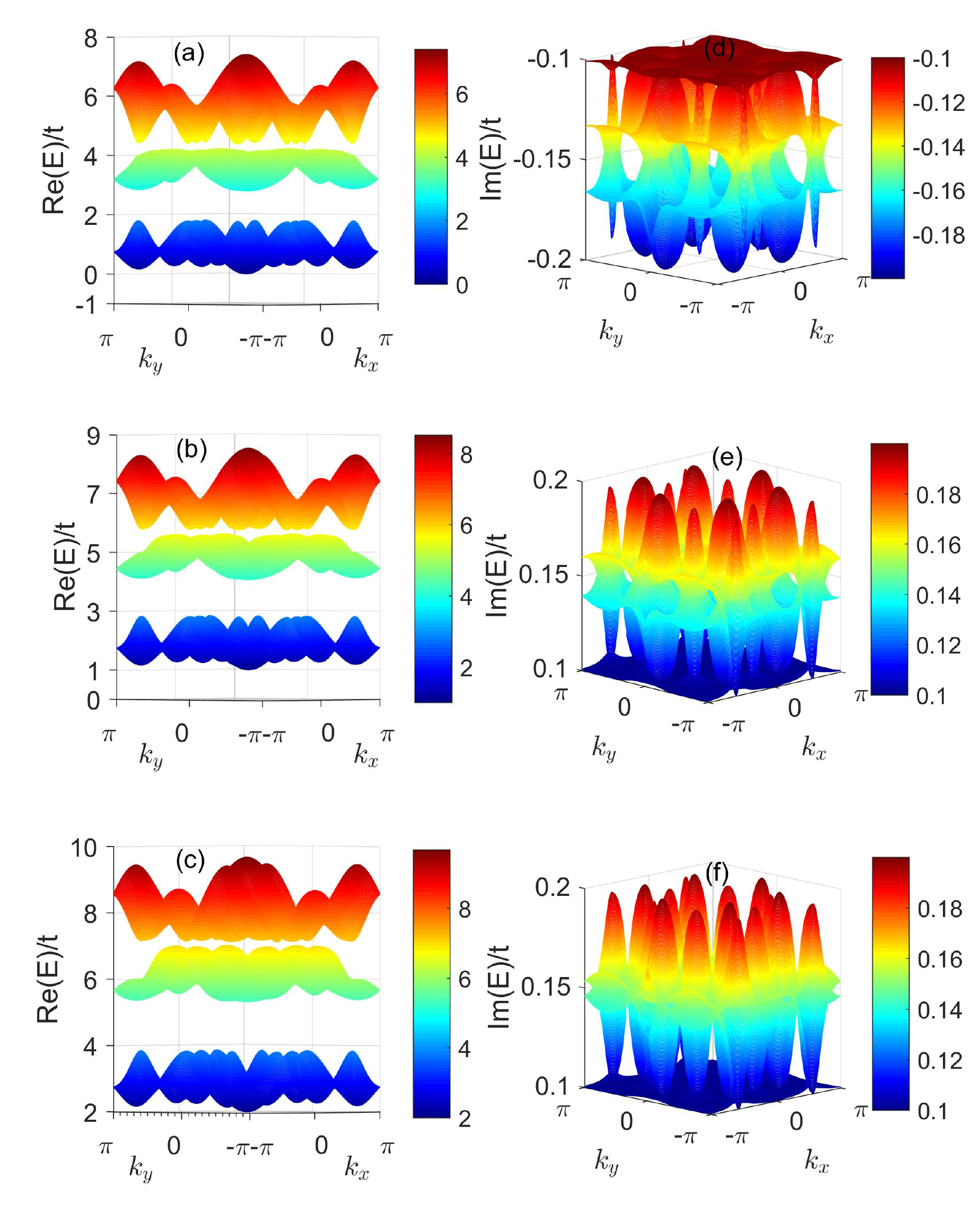}
\caption{ Energy spectra at the chosen parameter points in the imbalanced dissipation case. Top panel: energy spectra
at ($\eta_{a}, \Delta_{a}$), with the real part in (a) and the imaginary part in (d). Middle panel: energy spectra at ($\eta_{b}, \Delta_{b}$),
with the real part in (b) and the imaginary part in (e). Bottom panel: energy spectra at ($\eta_{c}, \Delta_{c}$), with the real
part in (c) and the imaginary part in (f). Intuitively, the bands in (a), (b) and (c) are isolated, whereas the energies
in the imaginary part are continuous. }
\label{f3}
\end{figure}

Figures.~\ref{f3}(a)-\ref{f3}(f) are the energy spectra at these chosen parameter points with Figs.~\ref{f3}(a), \ref{f3}(b), and \ref{f3}(c)
being the real spectra and Figs.~\ref{f3}(d), \ref{f3}(e), and \ref{f3}(f) being the corresponding imaginary spectra. We notice that all the real spectra have
separated bands, whereas the imaginary parts are continuous. This phenomenon means the topological property of energy bands in the imbalanced dissipation case is associated with the real gaps \cite{theory_1,theory_2}.

\begin{figure}[H]
\centering
\includegraphics[width=0.5\textwidth]{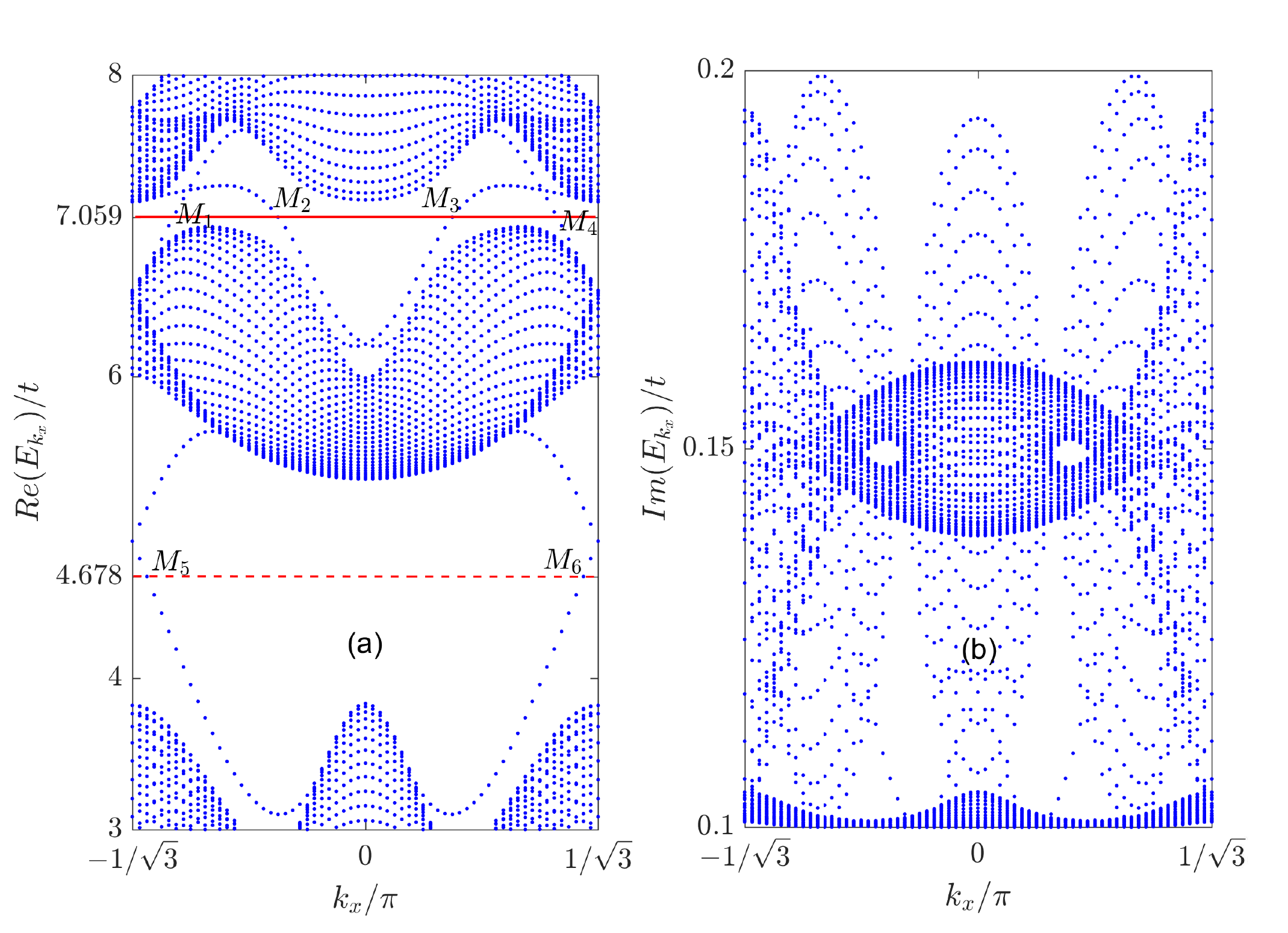}
\caption{Edge-state spectra with the zigzag edge at the chosen parameter point ($\eta_{c}, \Delta_{c}$) (parts
of the lower and higher Re($E_{k_{x}}$) are not shown). The bulk-edge correspondence is readily seen in the real spectrum.
(a) The real edge-state spectrum. Obviously, there are two pairs of edge modes within the upper bulk gap
and a pair of edge modes within the lower bulk gap, corresponding to $C_{\frac{2}{3}}=-2$ and $C_{\frac{1}{3}}=1$, respectively.
Four edge modes, labeled as $M_{1}$, $M_{4}$, $M_{2}$ and $M_{3}$ are chosen at Re($E_{k_{x}}$)=$7.059t$ (red solid line) and a
pair of edge modes, labeled as  $M_{5}$ and $M_{6}$ are chosen at Re($E_{k_{x}}$)=$4.678t$ (red dashed line).
(b) The imaginary part of the edge-state spectrum. Different form the real part in (a), this imaginary edge-state spectrum is
continuous, although there is a certain broadening. Meanwhile, no obvious bulk-edge correspondence phenomenon occurs in the
imaginary part. Other involved parameter is $N_{zigzag}=86$, and the number of discrete $k_{x}$ is 65.}
\label{f4}
\end{figure}

Further, we consider a semi-infinite cylindrical geometries with a zigzag edge to discuss the bulk-edge correspondence,
and the shape of geometry keeps the same as that in previous works \cite{dice_2}. To be precise, we take $N_{zigzag}=86$
which corresponds to the number of the lattice sites contained in the periodic repeating cell of the zigzag edge case. With the method of
partial Fourier transformation utilized in Ref. \cite{dice_2}, by choosing the parameter point ($\eta_{c}, \Delta_{c}$)
mentioned in this subsection before, we obtain the edge-state spectra as a function of $k_x$, shown in Figs.~\ref{f4}(a)
and \ref{f4}(b), respectively.

Figure~\ref{f4}(a) is the real edge-state spectrum in the zigzag edge case, where there are two pairs of edge modes within the upper
bulk gap and a pair of edge modes within the lower bulk gap, which can be reflected from the phase diagrams in Figs. \ref{f2}(a) and
\ref{f2}(b), showing the bulk-edge correspondence. Besides, from the edge-state spectra, it is shown that the edge states are protected
by the real gaps, which is consistent with the analyses of the energy spectra in Figs. \ref{f3}(c) and \ref{f3}(f). $M_{1}$ and $M_{4}$
and $M_{2}$ and $M_{3}$ are two pairs of edge modes chosen at Re($E_{k_{x}}$)=$7.059t$ (red solid line) and $M_{5}$ and $M_{6}$ are a pair
of edge modes chosen at Re($E_{k_{x}}$)=$4.678t$ (red dashed line). As for the imaginary edge-state spectrum in Fig. \ref{f4}(b), the energies
are continuous, and there is no obvioups bulk-edge correspondence in it.

\begin{figure}[H]
\centering
\includegraphics[width=0.5\textwidth]{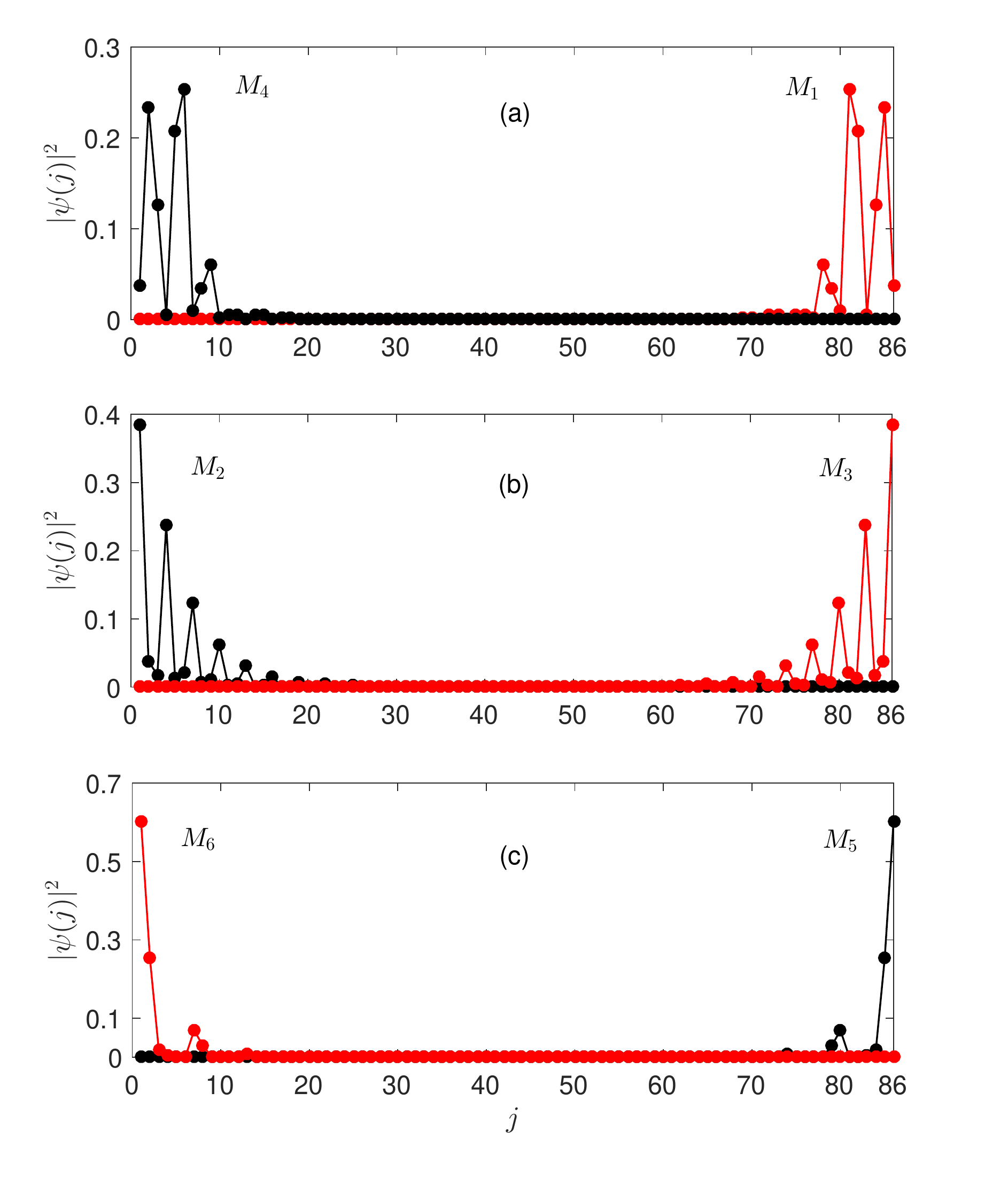}
\caption{The spatial density distributions of the six chosen edge modes $M_1$-$M_6$. Edge modes with a positive group velocity are shown in red,
whereas those with a negative group velocity are shown in black, presenting the chiral symmetry. $j$ is the site index. }
\label{f5}
\end{figure}

In order to further understand the bulk-edge correspondence, we plot the spatial density distributions of the six chosen edge modes
$M_{1}$-$M_{6}$ of the zigzag edge case, which are shown in Fig.~\ref{f5}. We conclude that the edge modes appear
in pairs with opposite momentums. Besides, the edge modes with positive group velocity are shown in red, whereas those with negative
group velocity are shown in black, presenting the chiral symmetry. We emphasize that the discussion of the spatial density distributions
of the edge modes in the armchair edge case \cite{dice_2} will yield similar results, which will not be analysed here any more. We find that the principle
of bulk-edge correspondence mentioned in Ref. \cite{bulk-edge_2} is still effective in the non-Hermitian case. Similar to its applications in
Ref. \cite{dice_2}, we focus on the edge modes localized at the $j=1$ side to discuss the relationship between the bulk Chern number and the edge
states. According to the phase diagram in Fig.~\ref{f2}(a), we know that the edge modes $M_{6}$ carries the Chern number $C=1$, and the edge
modes $M_{2}$ and $M_{4}$ with an opposite group velocity both of the Chern number $C=-1$. Hence, the Chern number of the middle band
is $C_{2}=-1+(-1)-1=-3$ which is self-consistent with the phase diagrams.

 \begin{figure}[H]
\centering
\includegraphics[width=0.5\textwidth]{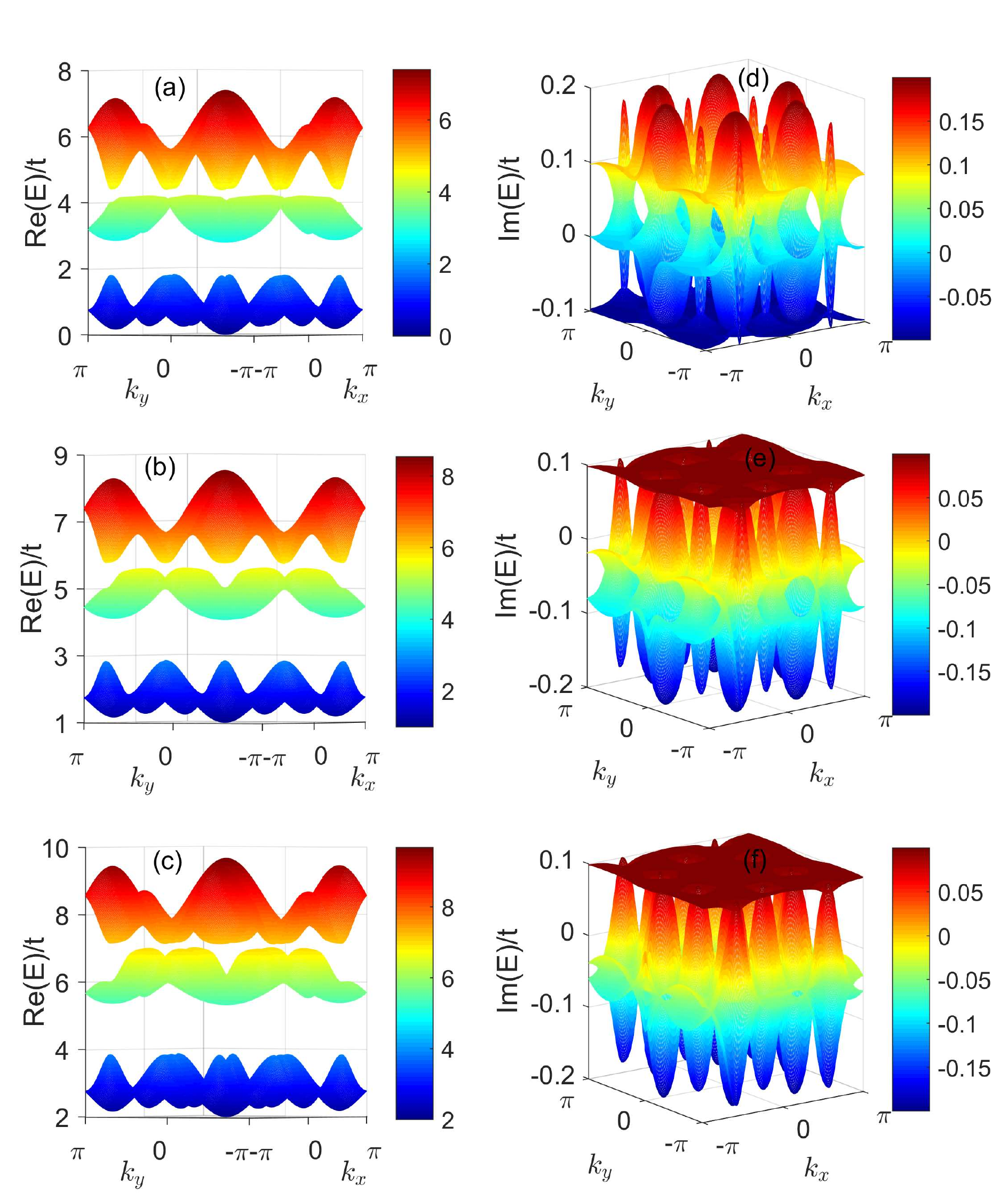}
\caption{Energy spectra at the chosen parameter points in the balanced dissipation case. Top panel: energy spectra
at ($\eta_{a}, \Delta_{a}$), with the real part in (a) and the imaginary part in (d). Middle panel: energy spectra at ($\eta_{b}, \Delta_{b}$),
with the real part in (b) and the imaginary part in (e). Bottom panel: energy spectra at ($\eta_{c}, \Delta_{c}$), with the real
part in (c) and the imaginary part in (f). Intuitively, the imaginary parts of energy spectra show an obvious
broadening and continuous band without separation. Instead, the bands in the real part are all clearly separated. }
\label{f6}
\end{figure}

\subsection{Balanced dissipation}
 A non-Hermitian system with balanced dissipations means a balanced gain and loss. For the purpose of studying topological properties
 of such a non-Hermitian with balanced dissipations, we specify $\chi=-1$ and $\eta_{1}=2\eta$ with $\eta \in [-0.1t, 0.1t]$. Then we substitute these
 parameters in Eq.~(\ref{eq7}) into the definition of Chern number and uncover that the obtained phase diagrams are consistent
 with the imbalanced dissipative system. Our numerical results show that the large Chern number is still robust in the balanced dissipation case.

 \begin{figure}[H]
 \centering
 \includegraphics[width=0.5\textwidth]{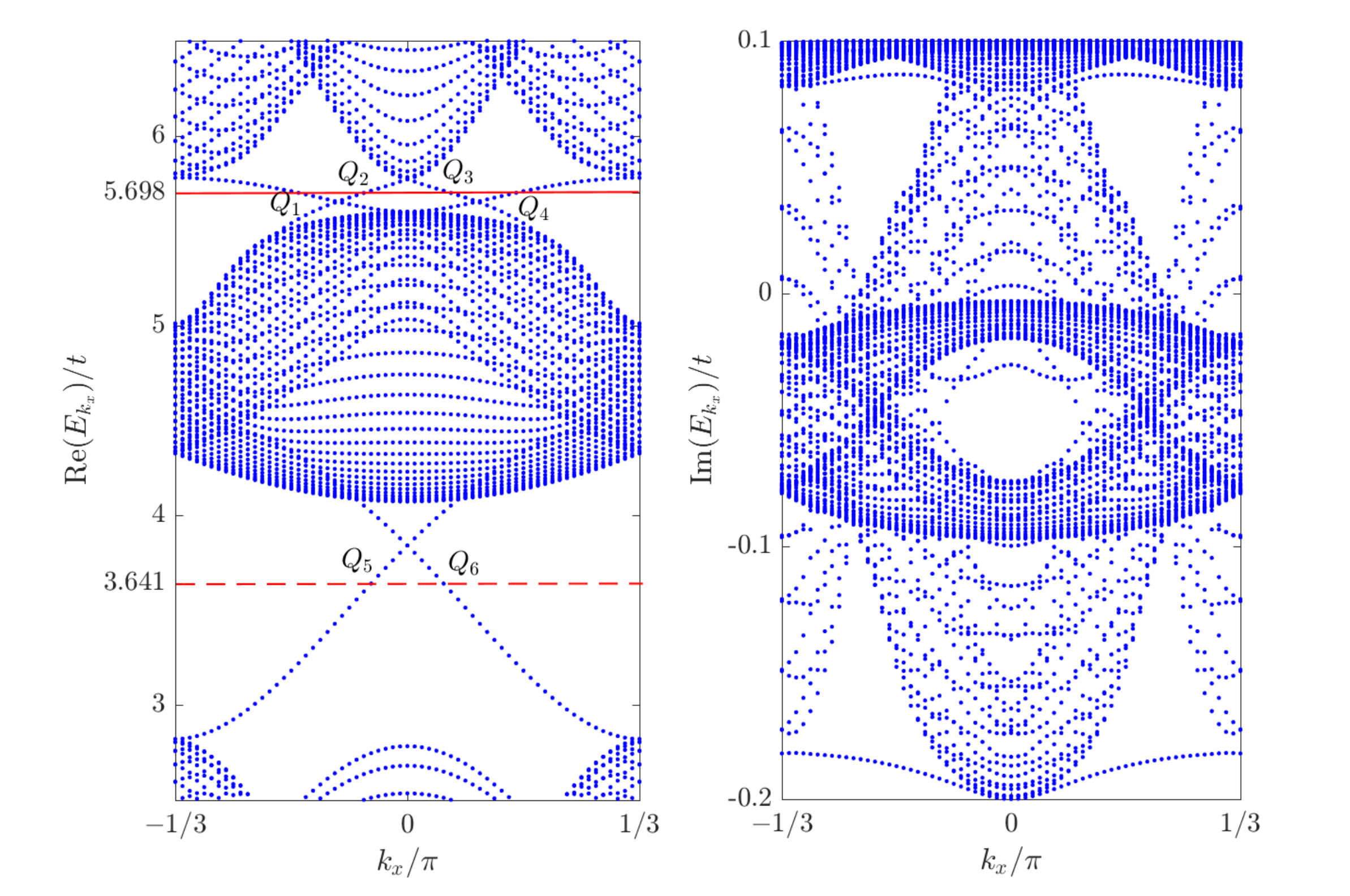}
 \caption{Edge-state spectra with an armchair edge. Left panel: Real edge-state spectra Re($E_{k_{x}}$) (Part of the lower and higher Re($E_{k_{x}}$) are not shown).
 Right panel: Imaginary edge-state spectra Im($E_{k_{x}}$). The bulk-edge correspondence can
 be readily seen in the real spectrum. As it shows, there are two pairs of edge modes within the upper bulk gap and a pair of edge modes
 within the lower bulk gap, corresponding to $C_{\frac{2}{3}}=-2$ and $C_{\frac{1}{3}}=1$, respectively. Four edge modes, labeled as $Q_{1}$, $Q_{4}$,
 $Q_{2}$ and $Q_{3}$ are chosen at Re($E_{k_{x}}$)=$5.698t$ (red solid line) and a pair of edge modes, labeled as  $Q_{5}$ and $Q_{6}$ are
 chosen at Re($E_{k_{x}}$)=$3.641t$ (red dashed line). In contrast, we cannot observe the bulk-edge correspondence in the imaginary spectrum
 clearly. Other involved parameter is $N_{armchair}=123$, and the number of discrete $k_{x}$ is 65.}
 \label{f7}
 \end{figure}

Similarly, energy spectra are plotted at three parameter points mentioned in the previous subsection, shown in Figs.~\ref{f6}(a)-\ref{f6}(f). The
 left panel shows the real spectra at three chosen parameter points, whereas the right panel corresponds to the imaginary parts. Although there is a
 certain broadening of the three spectra, the energies are continuous with no obvious band separation. On the contrary, there are distinct
 band gaps in the real parts. Similar to the imbalanced dissipation case, the topology of this system remains protected by the real gaps \cite{theory_1,theory_2}.
 Naturally, the bulk-edge correspondence of this balanced dissipative system can also be
 directly seen from real edge-state spectra. By choosing the parameter point ($\eta_{b}$, $\Delta_{b}$), we plot the associated edge-state spectra
 in the armchair edge case, shown in Figs.~\ref{f7}(a) and \ref{f7}(b), respectively.

Figure~\ref{f7}(a) is the real part of edge-state spectrum, in which there are distinct two pairs of edge modes within the upper bulk gap and a pair of
edge modes within the bottom bulk gap, corresponding to $C_{\frac{2}{3}}=-2$ and $C_{\frac{1}{3}}=1$, respectively. In contrast, there are no directly
visible bulk-edge correspondence in the imaginary spectrum (see Fig.~\ref{f7}(b)). In order to understand the bulk-edge correspondence adequately, four
edge modes, labeled as $Q_{1}$, $Q_{4}$, $Q_{2}$ and $Q_{3}$, are chosen at Re($E_{k_{x}}$)=$5.698t$ (red solid line), and a pair of edge modes,
labeled as  $Q_{5}$ and $Q_{6}$, are chosen at Re($E_{k_{x}}$)=$3.641t$ (red dashed line).

\begin{figure}[H]
\centering
\includegraphics[width=0.5\textwidth]{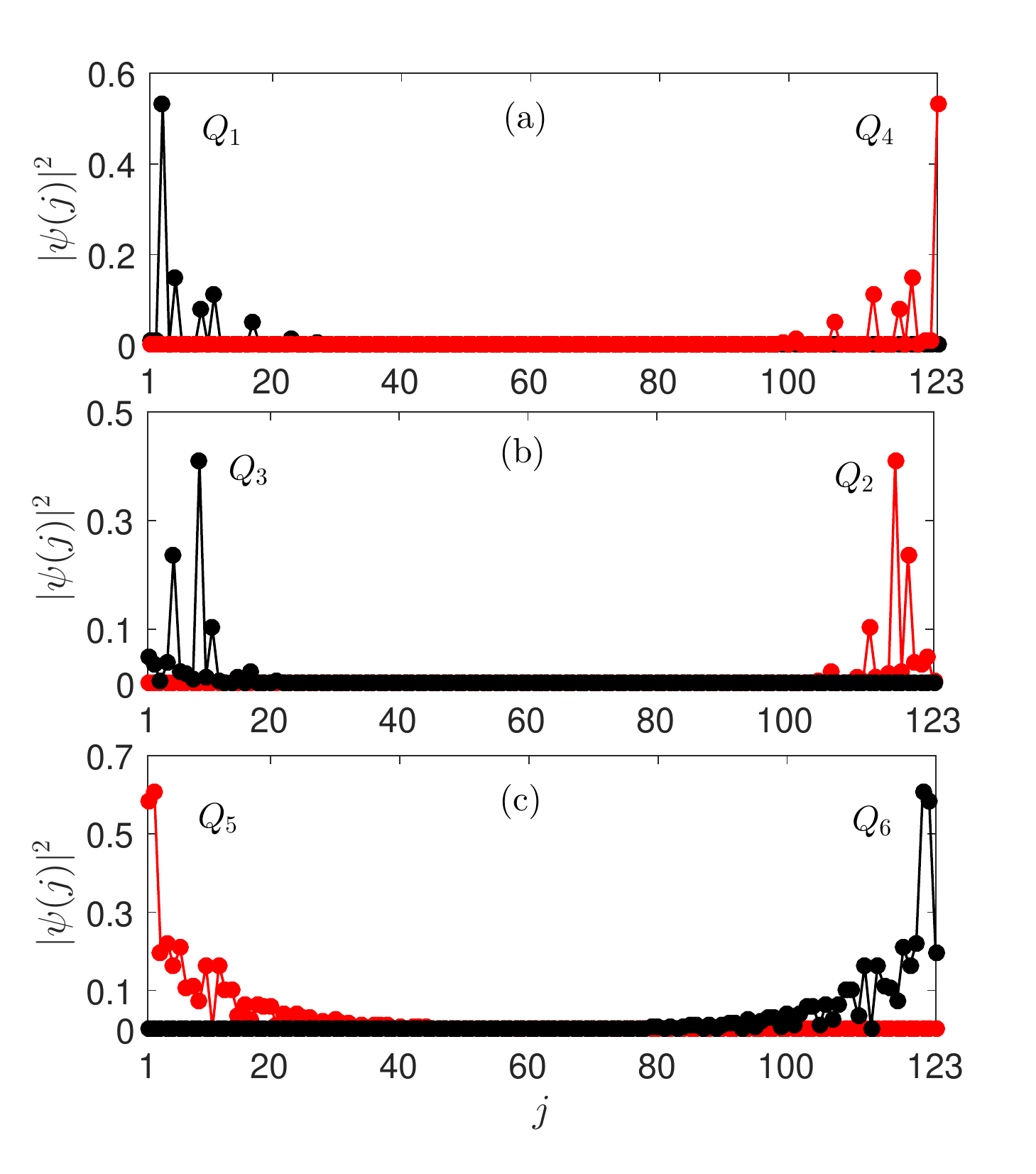}
\caption{The spatial density distributions of edge modes $Q_{1}$-$Q_{6}$. Edge modes with positive group velocity are shown in red,
whereas those with negative group velocity are shown in black, presenting the chiral symmetry. $j$ is the site index. }
\label{f8}
\end{figure}

Figure~\ref{f8} contains the spatial density distributions of these six chosen edge modes. Similarly, edge modes with positive group velocity
are shown in red, whereas those with negative group velocity are shown in black, presenting the chiral symmetry. Then an effective measure is taken
to analyze the relationship between the Chern numbers and corresponding edge states, namely, the principle of bulk-edge correspondence \cite{bulk-edge_2}.
We focus on the edge modes localized at the $j=N_{armchair}$ side. According to the phase diagram in Fig.~\ref{f2}(a), we know that the edge mode $Q_{6}$
carries the Chern number $C=C_{\frac{1}{3}}=1$, namely $C_{1}=1$. Therefore, the Chern number of edge modes $Q_{2}$ and $Q_{4}$, which are opposite
to the group velocity of $Q_{6}$, is $C=-1$. Consequently, the middle band has a large Chern number with $C_{2}=-1+(-1)-C_{1}=-3$. These
analyses are self-consistent with the phase diagrams and the real edge-state spectrum.

\section{\label{S5} Summary}
To sum up, a non-Hermitian system with an imbalanced dissipation and a balanced dissipation was studied. Our work suggests that the large Chern numbers
are robust against the non-Hermitian perturbations. Besides, from the energy spectra and the edge-state spectra, we have known that the associated
edge states are protected by the real bulk gaps. In addition, the relationship between the Chern numbers and the spatial density distributions of edge modes are
discussed by the principle of the bulk-edge correspondence, which is self-consistent with our phase diagrams and the edge-state spectra.

\section{Ackonwledge}
We thank the discussions with S. Chen and acknowledge the support from NSFC under
Grants No. 11835011 and No. 11774316.

\end{document}